%% file: soc_inside07.tex
\documentclass{format/acm_proc_article-sp}
\usepackage{framed}
\usepackage{listings}

\begin{document}

% display page numbers in the headings. Start with roman numerals %
%\pagestyle{headings}

%\setcounter{page}{79}

\pagenumbering{arabic}
\setpagenumber{25}

%%%%%%%%%%%%%%%%%%%%%%%%%%%%%%%%%%%%%%%%%%%%%%%%%%%%%%%%%%%%%%%%%%%%%%%%%%%

\title{On the adaptation of context-aware services}
%\subtitle{[Extended Abstract]
%\titlenote{A full version of this paper is available as
%\textit{Author's Guide to Preparing ACM SIG Proceedings Using
%\LaTeX$2_\epsilon$\ and BibTeX} at

\numberofauthors{1} %  in this sample file, there are a *total*
\author{
% 1st. author
%\alignauthor
Marco Autili, Vittorio Cortellessa, Paolo Di Benedetto, Paola
Inverardi\\
\\
       \affaddr{Dipartimento di Informatica }\\
       \affaddr{Università di L'Aquila}\\
       \affaddr{via Vetoio 1, L'Aquila, ITALY }\\
%       \email{marco.autili@di.univaq.it}
\email{\{marco.autili, cortelle, paolo.dibenedetto,
inverard\}@di.univaq.it}}

% There's nothing stopping you putting the seventh, eighth, etc.
% author on the opening page (as the 'third row') but we ask,
% for aesthetic reasons that you place these 'additional authors'
% in the \additional authors block, viz.
%\additionalauthors{Additional authors: John Smith (The Th{\o}rv{\"a}ld Group,
%email: {\texttt{jsmith@affiliation.org}}) and Julius P.~Kumquat
%(The Kumquat Consortium, email: {\texttt{jpkumquat@consortium.net}}).}

\date{17 September 2007}
% Just remember to make sure that the TOTAL number of authors
% is the number that will appear on the first page PLUS the
% number that will appear in the \additionalauthors section.

\maketitle
%%%%%%%%%%%%%%%%%%%%%%%%%%%%%%%%%%%%%%%%%%%%%%%%%%%%%%%%%%%%%%%%%%%%%%%%%%%
\input{abstract}
\input{introduction}
\input{context}
\input{developing}
\input{chameleon}
\input{deploying}
\input{conclusions}

%%%%%%%%%%%%%%%%%%%%%%%%%%%%%%%%%%%%%%%%%%%%%%%%%%%%%%%%%%%%%%%%%%%%%%%%%%%

\bibliographystyle{abbrv}
\nocite{}
\bibliography{soc@inside07}

\balancecolumns
\end{document}

%% file: abstract.tex
\begin{abstract}

Ubiquitous networking empowered by Beyond 3G 
networking makes it possible for mobile users to access networked
software services across heterogeneous infrastructures by
resource-constrained devices.
Heterogeneity and device limitedness creates serious
problems for the development and deployment of mobile services that are able to run
properly on the execution context and are able to ensures that users experience the ``best'' Quality of Service possible according to their needs and specific contexts of use.
To face these problems the concept of adaptable service is increasingly emerging in the software community.
In this paper we describe how CHAMELEON, a declarative framework
 for tailoring adaptable services, is used within the IST PLASTIC project whose goal is the rapid and easy
development/deployment of self-adapting services for B3G
networks.
%We introduce the notion of
%requested and offered Service Level Specification to address the (extra-functional) preferences of the user that
%will be used to establish the Service Level Agreement between the service consumer and
%the service provider. 
\end{abstract}

%%% Local Variables:
%%% mode: latex
%%% TeX-master: "ffaa"
%%% End:

%% file: introduction.tex
\section{Introduction}

Software pervades our life, at work and at home, spanning from
business to entertainment. We increasingly expect it to be
dependable and usable, despite of our own mobility, changing context
and needs. Ubiquitous networking empowered by Beyond 3G (B3G)
networking makes it possible for mobile users to access networked
software services across heterogeneous infrastructures by
(resource-constrained) devices, characterized by their limitedness
(e.g., smart phones, PDAs, etc.). Heterogeneity and limitedness
poses numerous challenges, among which we mention: developing
services that can be easily deployed on a wide range of evolving
infrastructures, from networks of devices to stand-alone wireless
resource-constrained hand-held devices; making services
resource-aware so that they can benefit from networked resources and
related services; and ensuring that users meet their
extra-functional requirements by experiencing the ``best'' Quality
of Service (QoS) possible according to their needs and specific
contexts of use.

Moreover, the extreme heterogeneity of mobile
terminals (e.g., processor, memory, display, I/O capabilities,
available radio interface, etc.) creates serious
problems for the development of mobile applications able to run
properly on a large heterogeneity of devices. To face these problems the concept of adaptable service is increasingly emerging in the software community. However, supporting the
development, deployment and execution of such adaptable services raises numerous
challenges that involve models, methods and tools.
Integrated solutions to these challenges are the main targets of the
IST PLASTIC project, whose goal is the rapid and easy
development/deployment of self-adapting services for B3G
networks~\cite{DoW}.

In this paper we briefly introduce the PLASTIC development process
model that relies on model-based solutions to build self-adaptable
context-aware services. Targeting adaptive services, this
development process focuses on the concept of context of use and
related Service Level Agreement (SLA). We introduce the notion of
\emph{requested Service Level Specification} (SLS) and \emph{offered
SLS} to address the (extra-functional) preferences of the user that
will be used to establish the SLA between the service consumer and
the service provider. The SLA is an
entity modeling the conditions on the QoS accepted by both the
service consumer and the service provider. In
~\cite{Lamanna:2003,Skene:2004} a language to precisely specify SLA
has been proposed. SLA represents a kind of contract that is
influenced by the service request requirements, the service
description and the context where the service has to be provided.
When a new service request is formulated, the PLASTIC platform has
to negotiate the QoS on the basis of the service
request, the context the service has to be provided and the service
descriptions of similar services already available by some
providers. The contractual procedure may terminate either with an agreement
about QoS of the service from the consumer and the provider, or with
no agreement.

In PLASTIC, adaptation is tackled at discovery time when the service
request is matched with a service provision. Due to the
heterogeneous nature of B3G environments, the service discovery
solution for PLASTIC provides mechanisms for supporting
dynamicity (service mobility, dynamic context information, dynamic
adaptation, etc.). Thus
the platform needs to dynamically discover services able to
correctly run on such devices and to guarantee the desired service's
quality expressed within the SLS requested by the user. This require ability to
reason on programs and environments in terms of the resources
they need and offer, respectively, and the ability to suitably adapt
the application to the environment that will host it.

In this setting, this paper focuses on the usage of a declarative
framework (called CHAMELEON) for tailoring adaptable services within
PLASTIC. We briefly present our Java based implementation of the
framework that is a refined/modified version of the original
framework whose foundations are presented in
\cite{Inverardi02,Inverardi04,Mancinelli07}. The approach makes use
of a \emph{light} extension of the Java language that is at the
basis of a declarative techniques to support the adaptation process,
the development and deploying of adaptable service applications. We
then discuss how services are discovered, accessed and deployed
within our framework. By leveraging this approach we are able to
perform a quantitative resource-oriented analysis of Java
applications. This analysis is relevant in the context of adaptable
application because it allows the framework to decide what
adaptation alternatives has to be chosen before the actual
deployment and execution.

For sake of space, we cannot address all the recent
related works in the wide domain of PLASTIC project, thus in the following we provide only some major references.
Current (web-)service development technologies, e.g.
~\cite{webtools,bpel,a-muse,wsdl,PDCS05} (just to cite some),
address only the functional design of complex services, that is they
do not take into account the extra-functional aspects (e.g., QoS
requirements) and the context-awareness. Our process borrows
concepts from these well assessed technologies and builds on them to
make QoS issues clearly emerging in the service development, as well
as to take into account context-awareness of services for
self-adaptiveness purposes.

The paper is structured as follow. Section \ref{sec:SettingTheContext} sets the ``context'' and Section \ref{sec:DevelopingContext-awareServices} introduces the PLASTIC development process model. Section \ref{sec:Chameleon} presents the framework CHAMELEON and the development environment it is based on. In Section \ref{sec:ServiceDiscoveryAccessAndDeployment} the PLASTIC service provision is discussed in term of service discovery, access, and deployment and finally Section \ref{conclusion} concludes and argues future work.

%% file: context.tex
\section{Setting the ``Context''}\label{sec:SettingTheContext}

\emph{Context awareness} and \emph{adaptation} have become two
important aspects in the development of service applications suited
to be executed in such an environment. In fact, as pointed out
in~\cite{Softure06,Inverardi06}, while delivering services,
applications need to be \emph{aware of} and \emph{adaptive to} the
context that is the combination of user-centric data (e.g.,
information of interest for the user according to his/her current
circumstance) and resource/computer-centric data (e.g., resource
constraints and conditions of the user device and network).

\emph{Context awareness} identifies the capability of being aware of
the user needs and of the resources offered by an execution
environment, in order to decide whether that environment is suited
to receive and execute the application in such a way that end-users
expectations are satisfied. \emph{Adaptation} identifies the
capability of changing the application in order to comply with the
current context conditions. In order to perform an adaptation it is
essential to provide an actual way to model the characteristics of
the application itself, of the heterogeneous infrastructures and of
the execution environment including the \emph{end-user degree of
satisfaction} (depending on requested and offered SLS). Thus, while
delivering services, it is useful to be able to reason about the
\emph{resources demanded} by an application (and its possible
adaptation alternatives) and the ones \emph{supplied} by the hosting
environment.

It is worthwhile stressing that although a change of context is
measured in terms of availability of resources, that is in
quantitative terms, an application can only be adapted by changing
its behavior - i.e., its functional/qualitative specification. In
particular, (Physical) Mobility allows a user to move out of his
proper context, traveling across different contexts. To our purposes
the difference among contexts is determined in terms of available
resources like connectivity, energy, software, etc. However other
dimensions of contexts can exist relevant to the user, system and
physical domains, which are the main context domains identified in
the literature \cite{schilit94contextaware}. In the software
development practice when building a system the context is
determined and it is part of the (extra-functional) requirements
(operational, social, organizational constraints). If context
changes, requirements change therefore the system needs to change.
Context changes occur due to physical mobility, thus while the
system is in operation. This means that if the system needs to
change this should happen dynamically.

%The context can be always the same during service lifetime, but can also change due to user
%physical mobility. This imply that, to respect functional and/or quantitative requirements, the service has to dynamically adapt .
%% This
%%adaptation can be achieved changing the structure and/or the
%%behavior of the system. For the structure, components can get in and
%%out, new connectors can be added and removed. For the behavior
%%components can change their functionality and connectors can change
%%their interaction protocols.
%Note that when is the behavior to
%change, mutations in the quantitative characteristics of the context
%are reflected to functional/qualitative mutations in the service.
In this setting, two different types of approach to the construction of adaptable
applications can be considered: \emph{self-contained} applications
that embody the adaptation logic as a part of the application
itself and, therefore, can handle dynamic changes in the environment by reacting to them at
runtime; \emph{tailored} applications that are the result of an
adaptation process, which has been previously applied on a generic
version of the application.
Self-contained adaptable applications
are inherently dynamic in their nature but suffer the pay-off of the inevitable overhead imposed by the
adaptation code. On the contrary, tailored adapted applications have a lighter code that make them suitable also for limited device, but are dynamic only with respect to the environment at deployment time, while remain static with respect to the actual
execution, i.e. they cannot adapt to runtime changes in the
execution environment.

Natively, the framework CHAMELEON is for tailored applications. However in Section \ref{conclusion} we argue how it can be extended toward a compromise between self contained and tailored service applications aiming at a more dynamic adaptation for limited device.

%approach in PLASTIC domain for dealing
%buttare qualcosa del tipo: As B3G network are characterized by limited device, the approach considered in this paper is typically tailored. In the conclusion, we will focus on an approach that represents a compromise between the two, aiming in bringing dynamic adaption in the world of limited device.

%The context represents the environment where the service has to be
%executed. From a software point of view it can be synthesized as the
%logical and physical resources available in the execution
%environment. These resources include hardware and software
%characteristic of the devices, network connectivity and reachable
%software services.

%% file: developing.tex
\section{Developing context-aware services}\label{sec:DevelopingContext-awareServices}

In this section we briefly introduce the PLASTIC development process
model that relies on model-based solutions to build self-adapting
context-aware services.

\begin{figure}[h]
  \centering
  \begin{framed}
    \includegraphics[width=8cm]{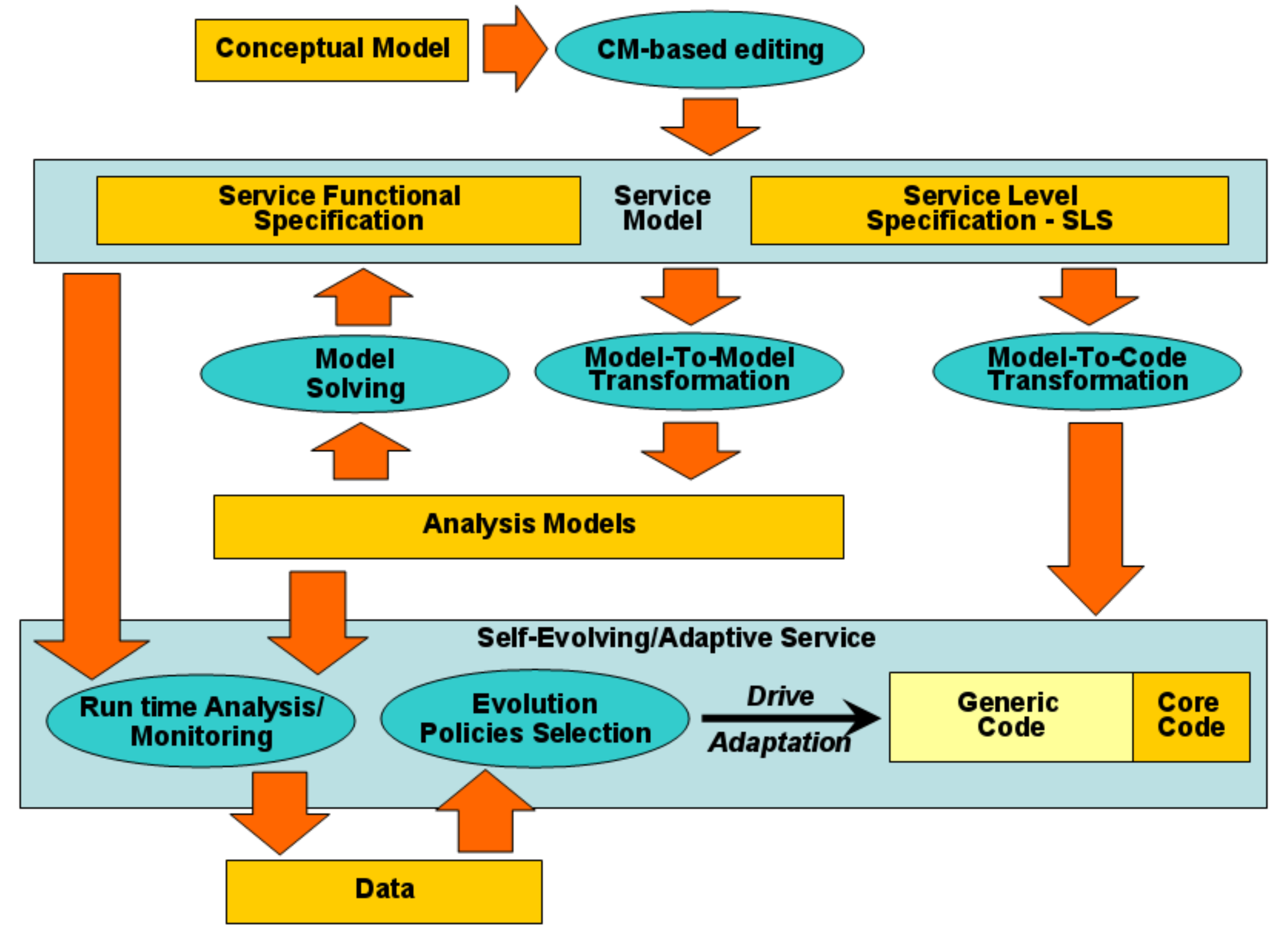}
  \end{framed}
  \caption{Plastic Development Process}
  \label{fig:PlasticDevelopmentProcess}
\end{figure}

By referring to Figure \ref{fig:PlasticDevelopmentProcess}, the
whole development process starts by taking into account the PLASTIC
Conceptual Model \cite{Deliverable2.1,Deliverable2.2} that is to say
a reference model for service-oriented B3G applications, which
formalizes the concepts needed to realize context-aware adaptable
applications for B3G networks. Within the PLASTIC Conceptual Model,
we specialized the concept \emph{context} in \emph{device context}
(provider and consumer side) and \emph{network context}. The former
supports the modeling of the possible devices, in terms of their
characteristics (e.g., screen resolution, CPU frequency, memory
size, etc.), with respect to which a service can perform adaptation.
The latter supports the modeling of the mobile nature of a PLASTIC
service deployed over different hosts/devices. Each PLASTIC enabled
device is connected to network(s) of type allowed by B3G open
wireless environment: all network characteristics are modeled as
network related context information (network context). These
information (retrieved by the PLASTIC middleware) together with its
device information (device context) realize the context-awareness of
a PLASTIC service being modeled and drives its adaptation.

In PLASTIC all the development process activities will originate
from the conceptual model and exploit as much as possible
model-to-model and model-to-code transformations. By taking into
account the conceptual model, the service model is specified in
terms of its functional specification and its SLS. The former
describes behavioral aspects of the modeled service, the latter its
QoS characteristics. Model-to-model transformation is performed in
order to derive models for different kinds of analysis. Some models,
e.g., stochastic models and behavioral models, are used at
development-time to refine/validate the service model (see the loop
``Service Model -> Model-To-Model Trans. -> Analysis Models -> Model
Solving -> Service Model'' shown in Figure
\ref{fig:PlasticDevelopmentProcess}). Some models (not necessarily
different from the previous one) will be made available at
deployment- and run-time to allow the adaptation of the service to
the execution context and service online validation, respectively
(see the box ``Self-Evolving/Adaptive Service'' shown in Figure
\ref{fig:PlasticDevelopmentProcess}). For a detailed description of
how the PLASTIC development process model has been instantiated we
refer to \cite{ICSOC07} where we describe how the service models are
specified (in terms of its functional specification and its SLS) and
how the model-to-model and model-to-code transformations are
performed.

We like to remark that one of the main novelties of the PLASTIC
process model is to consider SLS as part of a Service Model, as
opposite to existing approaches where SLS consists, in best cases,
in additional annotations reported on a (service) functional model.
This peculiar characteristic of our process brings several
advantages: (i) as the whole service model is driven by the
conceptual model, few errors can be introduced in the functional and
extra-functional specification of a service; (ii) SLS embedded
within a service model better supports the model-to-model
transformations towards analysis models and, on the way back, better
supports the feedback of the analysis; (iii) in the path to code
generation, the SLS will drive the adaptation strategies.
Model-To-Code transformation (right-hand side of Figure
\ref{fig:PlasticDevelopmentProcess}) is used to build both the core
and the adaptive code of the service. The core code is the frozen
portion of the developed self-evolving/ adaptive service. The
adaptive one is a ``generic code''.

A generic code embodies a certain degree of variability that makes
the code capable to evolve. This code portion is evolving in the
sense that, basing on contextual information and possible changes of
the user needs, the variability can be solved hence leading to a set
of alternatives. A particular alternative might be suitable for a
particular execution context and specified user needs. It can be
selected by exploiting the analysis models available at run-time and
the service capabilities performing both the ``Run time Analysis/SLA
Monitoring'' and the evolution policies selection. When a service is
invoked, the run-time analysis is performed (on the available
models) and, basing on the analysis results, a new alternative might be selected among the available ones.

The code is written by using the extended version of the Java
language used by the Development Environment of CHAMELEON framework
we are going to introduce.

%% file: chameleon.tex
\section{Chameleon}\label{sec:Chameleon}

\begin{figure}[h]
  \centering
  \begin{framed}
    \includegraphics[width=8cm]{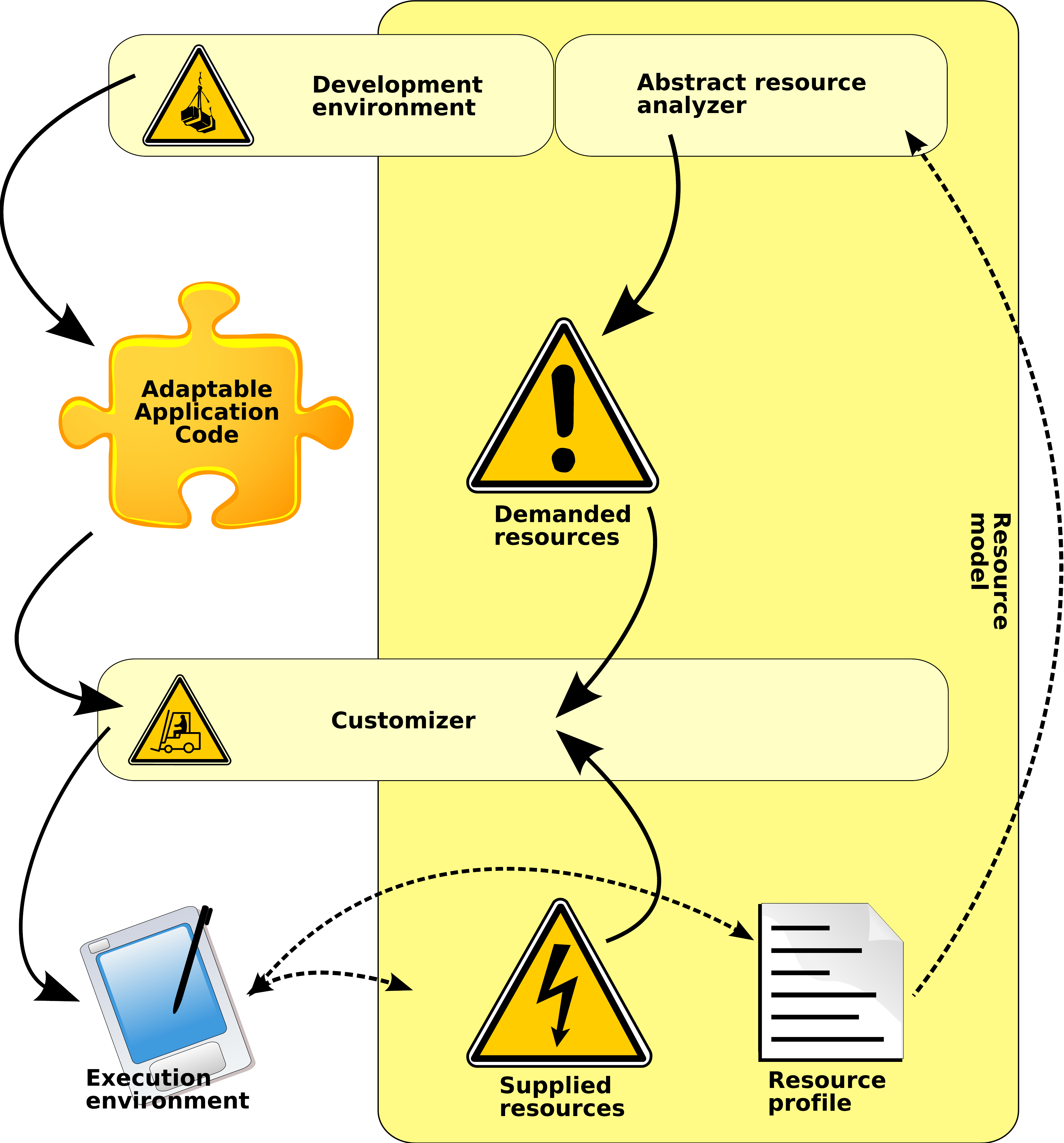}
  \end{framed}
  \caption{Chameleon Architecture}
  \label{fig:ChameleonArchitecture}
\end{figure}

The framework CHAMELEON aims at developing and deploying (Java)
adaptable service application. It supports the development of services
that are generic and can be correctly adapted with respect to a
dynamically provided context, which is characterized in terms of
available (hardware or software) resources, each one with its own
characteristics. To attack this problem we use a declarative and
deductive approach that enables the construction of a generic
adaptable service code and its correct adaptation with respect
to a given execution context \cite{Inverardi02, Inverardi04,
Mancinelli07}. Figure \ref{fig:ChameleonArchitecture} shows the
components of the framework's architecture.

%\begin{itemize}
%   \item
The \emph{Development Environment} is a standard Java development
environment that provides developers with a means for easily
specifying, in a flexible and declarative way, how the service can
be adapted. Considering methods as the smallest building blocks that
can be adapted in the service code, the model uses some ad-hoc
extensions to the reference language, i.e.,
Java~\cite{Java,Eckel03}, to express adaptation. Specifically, the
standard Java syntax is enriched by dedicated key-word and
annotations that permit to specify the following elements:
\emph{adaptable classes} that are classes that contain one or more
\emph{adaptable methods}; {\em adaptable methods} that are the
entry-points for a behavior that can be adapted; finally {\em
adaptation alternatives} that specify how one or more
\emph{adaptable methods} can actually be adapted. That is, the
extended Java syntax is used for specifying generic service code. In
order to be conservative with respect to the existing tools, we
developed a Java implementation of a preprocessor that takes as
input generic service code and translates it into a standard Java
program that can be processed by traditional IDEs and compiled by
using traditional Java compilers.

\begin{figure}[h]
\begin{framed}
{%\linespread{0.2}
\begin{lstlisting}[language=java,basicstyle=\small\sffamily,morekeywords={tag,adaptable,alternative,adapts}]
public adaptable class Connection {
  ...
  public adaptable void send();
  public adaptable void connect();
  ...
}

alternative Bluetooth adapts Connection {

  public send() {
  //send message using Bluetooth adapter
  }

  public connect() {
  //connect by Bluetooth
  }

}

alternative Wifi adapts Connection {

  public send() {
  //send message using WiFi adapter
  }

  public connect() {
  //connect by WiFi
  }

}
\end{lstlisting}
}
\end{framed}
  \caption{An adaptable class}
  \label{fig:adaptableClass}
\end{figure}

Figure~\ref{fig:adaptableClass} represent a chunk of a simple adaptable service that has been
written using the above extensions. Specifically, the
\emph{adaptable class} {\sffamily Connection} contains two
\emph{adaptable methods}: {\sffamily send} and {\sffamily connect} (see the non standard
key-word \textbf{{\sffamily adaptable}}). Note that, adaptable
methods do not have a definition in the adaptable class where they
are declared but they are defined within \emph{adaptation
alternatives} (see the keyword \textbf{{\sffamily alternative}}). In
general, it is possible to specify more than one alternative for a
given adaptable class provided that for each adaptable method there
exists at least one alternative that contains a definition for it.
The Connection class has two alternatives; one that connect and send messages using Bluetooth network adapter and the other one uses WiFi adapter.

%   \item

The \emph{Abstract Resource Analyzer} examines the service code
written in the Development Environment and extracts from it a
declarative description of its characteristics in terms of resource
demands. Actually, the analyzer is an implementation of an abstract
semantics that interprets the code with respect to a well defined
Resource Model, and extracts the information according to that
model.

%   \item

The \emph{Customizer} takes care of exploring the space of all the
possible adaptation alternatives and carries out the actual
adaptation before the deployment in the target environment for
execution. This step produces a standard Java service code.
%   \item

The \emph{Execution Environment} can be any device, equipped with a
standard java virtual machine, that will host the execution of the
service. Typically the Execution Environment will be provided by
Personal Digital Assistants (PDA), mobile phones, smart phones, etc.
From this point of view, the Execution Environment is not strictly
part of the framework we are presenting here. However it must be
characterized by a declarative description of the resources it
provides (i.e., the resource supply) that are retrieved by an
additional software component deployed on the Execution Environment
itself.
%   \item

The \emph{Resource Model} is a formal model in which it is possible to clearly specify the characteristics
with respect to resource aspects of the services and the environments that are handled by the
framework. The Resource Model, moreover, enables the framework to reason on the adaptation alternatives and allows it to choose the "best" one depending on several factors. It is spread throughout the whole framework.
%\end{itemize}

%% file: deploying.tex
\section{Service Discovery, Access and Deployment}\label{sec:ServiceDiscoveryAccessAndDeployment}

The PLASTIC service provision/consumption will be based on the Web
Services (WS) technology that provides a standard means for
interoperating between (distributed) software services by means of
the Web Services Interaction Pattern \cite{WSC}.

\begin{figure}[h]
  \centering
  \begin{framed}
    \includegraphics[width=8cm]{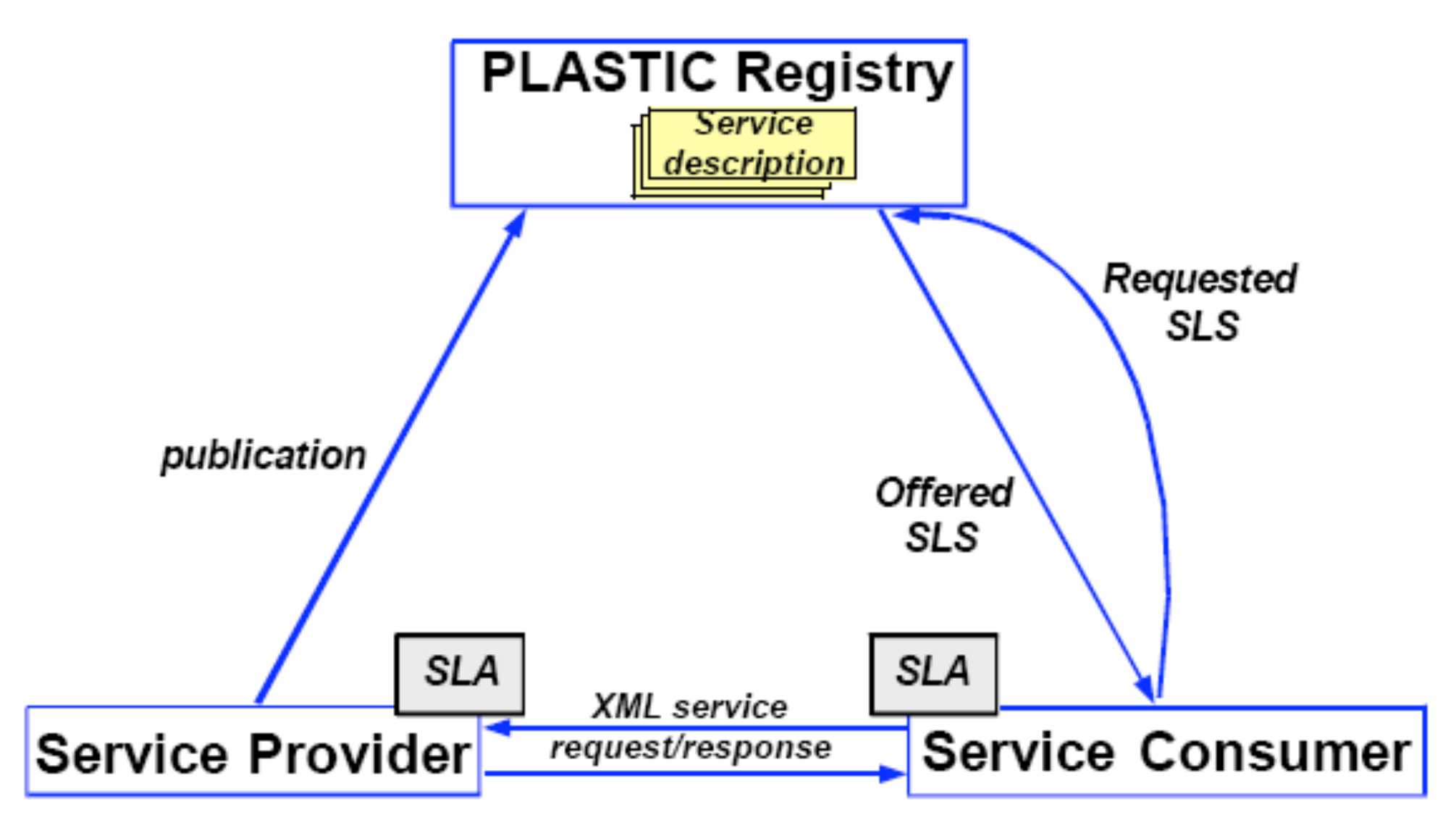}
  \end{framed}
  \caption{PLASTIC Services Interaction Pattern}
  \label{fig:plasticWSDL}
\end{figure}

Within PLASTIC, the WS interaction pattern is slightly modified in
order to reach the SLA at the end of the discovery phase.
Considering the PLASTIC adaptation empowered by CHAMELEON, the
discovery process has to take into account the user's QoS request
(i.e., the requested SLS) and the service SLSs (i.e., the offered
SLSs) in order to deliver a suitably adapted consumer application
(i.e., the right alternative) that, deployed on the user device,
will properly run and will allow for the consumption of the service
satisfying the requested SLS. That is, accounting for the different
SLSs associated to the different alternatives, a matching procedure
is started trying to produce the SLA defining the QoS constraints
under which the service should operate (i.e., the contract).\\
Specifically the steps involved in the PLASTIC service provision and
consumption are the following (see Figure \ref{fig:plasticWSDL}):
%
%
%
%a possible scenario is the one of a user that wants to download a
%suitably adapted client application that allows for the consumption
%of a particular service under the requested QoS constraints
%

\begin{enumerate}

\item The service provider publishes into the PLASTIC Registry the service
description in terms of both functional specifications and
associated SLSs. In fact, as already said, each service can be
implemented by different adaptation alternatives (generated by the
Customizer), each one characterized by its own SLS. SLSs are
computed by the service provider, on the base of the resource
consumption of the alternative analyzed by the Abstract Resource
Analyzer. The native WSDL is extended to deal with the SLS
specifications and, differently from the already in place service
registries, the PLASTIC Registry is able to deal with these
additional information in order to choose the most suitable
alternative.

\item The service consumer queries the PLASTIC Registry for a specific
service functionality, additionally providing the device resource
supply and the requested SLS.

\item The PLASTIC Registry searches for an offered SLS that satisfies the
requested SLSs with the provided resource supply. Whenever there
exists an adaptation alternative that has associated a suitable
offered SLS, the SLA can be established. Then, part of the extended
WSDL (published by the service provider) is passed to the service
consumer so that the service, and in particular the suitable
alternative, can be located. If no suitable alternative is able to
directly and fully satisfy the requested requirements, negotiation
is necessary. The negotiation phase starts by proposing a set of
alternatives that are suitable according to the resource supply but
whose offered SLS does not fully match the required SLS. Hence, the
consumer can perform a new request and the process is reiterated
till an SLA is possibly reached.

\item If the previous phase is successful - i.e., the SLA has been reached
- the right alternative can be delivered and deployed on the
consumer device, and the service consumption can take place under
the QoS constraints.

\end{enumerate}

For instance, consider the service that permits to connect and send
messages whose meta-code is illustrated in
Figure~\ref{fig:adaptableClass}. The service code has two possible
adaptations. The provider publishes the service into the PLASTIC
Registry by associating to it two possible offered SLSs: $SLS_{bt} =
\{Speed=Low, Cost=Low\}$ and $SLS_{wf} = \{Speed=High, Cost=High\}$
associated to the Bluetooth and WiFi alternatives, respectively. Let
us suppose that a consumer searching for the service has a device
that has only the \emph{WiFi network adapter} as part of its
resource supply (i.e., the device is equipped by only the WiFi radio
interface). Moreover, he/she specifies a requested SLS
$SLS_{req}=\{Cost=Low\}$. The only alternative that suits the
resource supply is WiFi but $SLS_{wf}$ does not fully match
$SLS_{req}$. Thus, a negotiation is necessary and it starts by
informing the client that the PLASTIC platform can only offer the
service alternative associated to $SLS_{wf}$. If the consumer
accepts, the SLA is reached, the extended WSDL containing the
reference to the WiFi alternative is provided, the service
application is deployed on the consumer device, and the service
consumption can take place.

Note that, the described ``consumer-side'' scenario also applies
whenever a user wants to provide a service (and hence wants to play
the role of service provider). In this case the (set of) suitable
alternative(s) can be delivered to and then deployed on the user
device, and the service provision can take place.

%% file: conclusions.tex
\section{Discussion and Future Work}\label{conclusion}
In this paper we described how a declarative framework
 for tailoring adaptable services (called CHAMELEON) \cite{Inverardi02,Inverardi04,Mancinelli07} is used within IST PLASTIC project~\cite{DoW}.
This framework is at the basis of the PLASTIC process model \cite{Deliverable2.1,Deliverable2.2} for the development, deployment, and validation of adaptable software services targeted to mobile (resource-constrained) devices running on heterogeneous network infrastructures.

Right now, PLASTIC provides a limited form of adaptation. In fact
adaptation happens at the time the service request is matched with a
service provision. Thus the deployed service is customized with
respect to the context at deployment time but, at run time, it is
frozen with respect to evolution. For instance, in B3G scenarios a
typical problem is represented by the fact that users are movable
and physical mobility imply changes to the context. This means that
if the service has to continue respecting the reached SLA and
dependability, it needs to dynamically adapt.

As said before the variety of possible configuration of B3G network,
is so wide to think to deploy a self-adaptive service that will be suitable for
any possible context in which the user can move, also considering
the limitedness of devices. In the same time it will be difficult to
reach an SLA that is not too loose. To cope this problem, we can
fairly assume that the user, at the moment of service request, knows (at least a stochastic distribution of)
the mobility pattern he will follow during service usage. This will
introduce some amount of determinism that permits to identify the
successive finite contexts the user can find during the service
usage.

Assuming that the user at the moment of the service request,
specifies his mobility pattern, we can predictively evaluate the
impact that the identified mobility pattern have on the service
performance, using the methodology proposed in \cite{antinisca07} for
modeling performance of physically mobile systems.
The approach generates Layered Queuing Network (LQN) models from the
description of the software architecture of the application, of
the considered user mobility patterns, and of the context (hardware
plus software) the application meets during the user mobility. Then,
it evaluates the obtained models in order to predict the performance
indexes of interests. Indeed, the LQN generation algorithm derives a
set of LQN models, one for each system configuration identified in
the Physical Mobility description, and calculates performance
metrics that estimates the performance of the software system when a
user/device has one of the Physical Mobility behaviors described.
These metrics can be used at discovery phase to define the offered
SLS in presence of user mobility conform to the identified patterns.

Following this approach, an enhanced version of CHAMELEON would be
able to generate a service code that is a compromise between
self-contained and tailored adaptable code. The code would embed all
the adaptation alternatives necessary to preserve the offered SLS
associated to the specified mobility pattern. Moreover, the service
code would implement some dynamic adaptation logic which is able to
recognize context changes further switching among adaptation
alternatives. That is, the deployed service - though running on a
limited device - would be able to evolve at run-time for adapting
itself to the sensed context changes according to the set of
adaptation alternatives considered at deployment time.

\textbf{Acknowledgments.} This work has been partially supported by
the IST EU project PLASTIC (www.ist-plastic.org).

%
%In this way we would use an approach that is a mix
%between self-contained and tailored adaptable applications realizing
%a compromise that aims at deploying dynamically adaptable
%application still limiting the overhead imposed by the adaptation
%code.
%

%% file: soc_inside07.bbl
\begin{thebibliography}{10}

\bibitem{Softure06}
{A. Bertolino and W. Emmerich and P. Inverardi and V. Issarny}.
\newblock {Softure: Adaptable, Reliable and Performing Software for the
  Future}.
\newblock {\em {Future Research Challenges for Software and Services (FRCSS)}},
  {2006}.

\bibitem{ICSOC07}
M.~Autili, L.~Berardinelli, V.~Cortellessa, A.~D. Marco, D.~D. Ruscio,
  P.~Inverardi, and M.~Tivoli.
\newblock A development process for self-adapting service oriented
  applications.
\newblock In {\em Proceedings of the International Conference on Service
  Oriented Computing (ICSOC)}, Vienna, Austria, 2007. To appear.

\bibitem{Lamanna:2003}
{D. Lamanna, J. Skene, W. Emmerich}.
\newblock Slang: A language for defining service level agreements.
\newblock In {\em Proc. FTDCS}, pages 100--107, San Juan, Puerto Rico, {2003}.

\bibitem{Eckel03}
B.~Eckel.
\newblock {\em Thinking in Java}.
\newblock Prentice Hall Professional Technical Reference, 2003.

\bibitem{webtools}
{E}clipse.org.
\newblock {E}clipse {W}eb {S}tandard {T}ools.
\newblock http://www.eclipse.org/webtools.

\bibitem{WSC}
H.~K. G.~Alonso, F.~Casati and V.~Machiraju.
\newblock {\em Web Services: Concepts, Architectures and Applications}.
\newblock Springer-Verlag Berlin Heidelberg, 2004.

\bibitem{bpel}
IBM.
\newblock {BPEL4WS}, {B}usiness {P}rocess {E}xecution {L}anguage for {W}eb
  {S}ervices, version 1.1, 2003.

\bibitem{Inverardi06}
P.~Inverardi.
\newblock Software of the future is the future of software?
\newblock In {\em Proceedings of the second symposium on Trustworthy Global
  Computing (TGC2006)}, Lucca, Italy, 2006. LNCS volume. To appear.

\bibitem{Inverardi02}
P.~Inverardi, F.~Mancinelli, and G.~Marinelli.
\newblock Correct deployment and adaptation of software applications on
  heterogenous (mobile) devices.
\newblock In {\em WOSS '02: Proceedings of the first workshop on Self-healing
  systems}, pages 108--110, New York, NY, USA, 2002. ACM Press.

\bibitem{Inverardi04}
P.~Inverardi, F.~Mancinelli, and M.~Nesi.
\newblock A declarative framework for adaptable applications in heterogeneous
  environments.
\newblock In {\em SAC '04: Proceedings of the 2004 ACM symposium on Applied
  computing}, pages 1177--1183, New York, NY, USA, 2004. ACM Press.

\bibitem{Skene:2004}
{J. Skene, D. Lamanna, W. Emmerich}.
\newblock Precise service level agreements.
\newblock In {\em Proc. of the 26th ICSE.}, pages 179--188, Edinburgh, UK,
  {May} {2004}.

\bibitem{Mancinelli07}
F.~Mancinelli and P.~Inverardi.
\newblock Quantitative resource-oriented analysis of java (adaptable)
  applications.
\newblock In {\em WOSP '07: Proceedings of the 6th international workshop on
  Software and performance}, pages 15--25, New York, NY, USA, 2007. ACM Press.

\bibitem{antinisca07}
A.~D. Marco and C.~Mascolo.
\newblock Performance analysis and prediction of physically mobile systems.
\newblock In {\em WOSP '07: Proceedings of the 6th international workshop on
  Software and performance}, pages 129--132, New York, NY, USA, 2007. ACM
  Press.

\bibitem{Deliverable2.1}
{PLASTIC IST STREP Project}.
\newblock {Deliverable D2.1: SLA language and analysis techniques for adaptable
  and resource-aware components}.
\newblock
  http://www-c.inria.fr/plastic/deliverables/plastic-d2\_1-finalpdf.pdf/download.

\bibitem{Deliverable2.2}
{PLASTIC IST STREP Project}.
\newblock {Deliverable D2.2: Graphical design language and tools for
  resource-aware adaptable components and services}.
\newblock
  http://www-c.inria.fr/plastic/deliverables/plastic-d2\_2-finalpdf.pdf/download.

\bibitem{a-muse}
A.-M. Project.
\newblock {Methodological Framework for Freeband Services Development}, 2004.
\newblock https://doc.telin.nl/dscgi/ds.py/Get/File-47390/.

\bibitem{DoW}
P.~Project.
\newblock {Description of Work}, 2005.
\newblock http://www.ist-plastic.org.

\bibitem{schilit94contextaware}
B.~Schilit, N.~Adams, and R.~Want.
\newblock Context-aware computing applications.
\newblock In {\em {IEEE} Workshop on Mobile Computing Systems and
  Applications}, Santa Cruz, CA, US, 1994.

\bibitem{Java}
{SUN Microsystems}.
\newblock The java language specification (third edition), 2005.
\newblock http://java.sun.com/docs/books/jls/index.html.

\bibitem{wsdl}
W3C.
\newblock {W}eb {S}ervice {D}efinition {L}anguage, http://www.w3.org/tr/wsdl,
  2001.

\bibitem{PDCS05}
H.~Yun, Y.~Kim, E.~Kim, and J.~Park.
\newblock {Web Services Development Process}.
\newblock In {\em PDCS}, 2005.

\end{thebibliography}
